\documentclass[sigconf]{acmart}

\AtBeginDocument{%
  \providecommand\BibTeX{{%
    \normalfont B\kern-0.5em{\scshape i\kern-0.25em b}\kern-0.8em\TeX}}}

\setcopyright{acmcopyright}
\copyrightyear{2023}
\acmYear{2023}
\acmDOI{10.1145/3544549.3585900}

\usepackage{caption}
\usepackage{subcaption}

\acmConference[CHI '23]{Make sure to enter the correct
  conference title from your rights confirmation emai}{April 23--28,
  2023}{Hamburg, Germany}
\acmBooktitle{Hamburg '23: ACM Transactions on Computer-Human Interaction,
 April 23--28,
  2023, Hamburg, Germany} 
\acmPrice{15.00}
\acmISBN{978-1-4503-XXXX-X/18/06}

\copyrightyear{2023}
\acmYear{2023}
\setcopyright{rightsretained}
\acmConference[CHI EA '23]{Extended Abstracts of the 2023 CHI Conference on
Human Factors in Computing Systems}{April 23--28, 2023}{Hamburg, Germany}
\acmBooktitle{Extended Abstracts of the 2023 CHI Conference on Human
Factors in Computing Systems (CHI EA '23), April 23--28, 2023, Hamburg,
Germany}\acmDOI{10.1145/3544549.3585900}
\acmISBN{978-1-4503-9422-2/23/04}

\begin{document}



\title{Experts prefer text but videos help novices: an analysis of the utility of multi-media content}

\author{Hayeong Song}
\affiliation{%
  \institution{Georgia Institute of Technology}
  \city{Atlanta}
  \country{USA}}
\email{hsong300@gatech.edu}

\author{Jennifer Healey}
\affiliation{%
  \institution{Adobe Research}
  \city{San Jose}
  \country{USA}}
\email{JeHealey@adobe.com}

\author{Alexa Siu}
\affiliation{%
  \institution{Adobe Research}
  \city{San Jose}
  \country{USA}}
\email{asiu@adobe.com}

\author{Curtis Wigington}
\affiliation{%
  \institution{Adobe Research}
  \city{San Jose}
  \country{USA}}
\email{wigingto@adobe.com}

\author{John Stasko}
\affiliation{%
  \institution{Georgia Institute of Technology}
  \city{San Jose}
  \country{USA}}
\email{stasko@cc.gatech.edu}

\renewcommand{\shortauthors}{Song et al.}

\begin{abstract}
  Multi-media increases engagement and is increasingly prevalent in online content including news, web blogs, and social media, however, it may not always be beneficial to users. To determine what types of media users actually wanted, we conducted an exploratory study where users got to choose their own media augmentation. Our findings showed that users desired different amounts and types of media depending on their familiarity with the content. To further investigate this difference, we created two versions of a media augmented document, one designed for novices and one designed for experts. We evaluated these prototypes in a two-way between-subject study with 48 participants and found that while multi-media enhanced novice readers' perception of usability (p = .0100) and helped them with reading time (p = .0427), time on task (p= .0156), comprehension (p = .0161), experts largely ignored multi-media and primarily utilized text.
\end{abstract}





\begin{teaserfigure}
  \includegraphics[width=\textwidth]{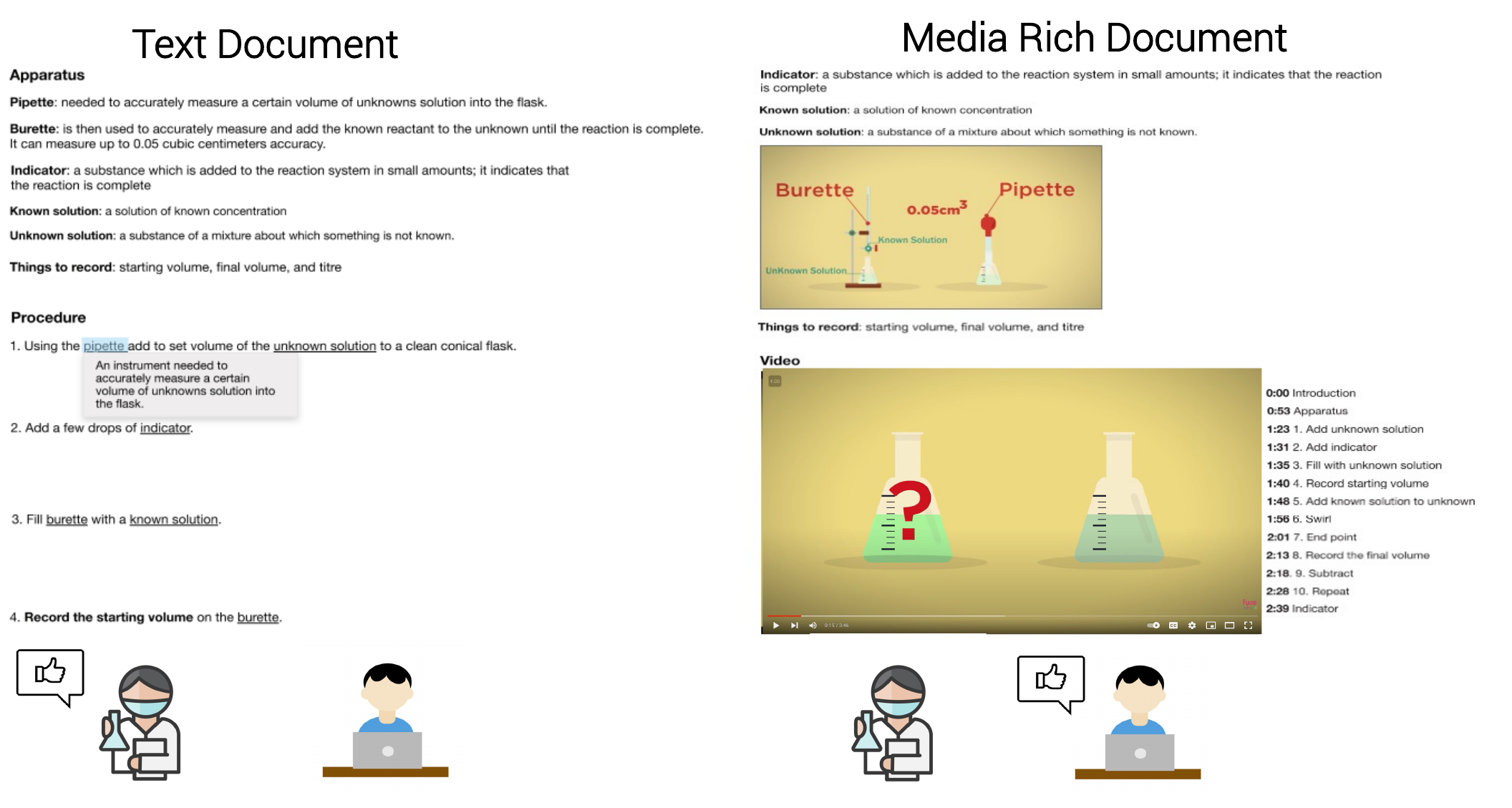}
  \caption{We designed two versions of the media-augmented documents for \emph{novices} and \emph{experts}. A \emph{text} document (all information in text) designed for \emph{experts} had in-line text augmentation and media \emph{rich} document (text, images, animated gifs, and a video) designed for \emph{novices} had multi-media content. We evaluated these prototypes in a two-way between-subject study with 48 participants and found that while multi-media enhanced novices' perception of usability and helped them with reading time, quiz time, comprehension, experts largely ignored multi-media and primarily utilized text.}
  \Description{Text document has all information in text with in-line text augmentation. Media rich document has text, image, animated gifs, and video. An expert is showing more positive sentiment toward the text document and a novice is showing more positive sentiment toward the media rich document.}
  \label{fig:teaser}
\end{teaserfigure}

\maketitle

\newcommand{\hs}[1]{\textbf{\textcolor{magenta}{#1}}}

\section{Introduction}


We believe that in the near future documents will naturally include multi-media content such as interactive elements, animations, and videos. Our belief is based on the observations that text documents continue to include more images and that information is increasingly shared through infographics, slides, and video presentations. On the internet, the multi-media content of digital documents such as newspapers, web blogs, and social media is continually increasing~\cite{Badger}. While it has been shown that multi-media content increases \emph{engagement}, for example, tweets with video are shared ten times more than tweets without video~\cite{twitter}, it has not been well studied whether or not multi-media content is truly \emph{desired} or \emph{helpful} to people seeking information. We hypothesize that both future documents will contain multi-media content and that digital documents will be able to customize the presentation of this content differently for different people. In this paper, we present our design process for motivating such customized reading experiences. 

To investigate how people might want a text document to be augmented, we conducted a formative study with ten participants. In the study, we gave people a plain text procedural document about chemistry titration lab and asked them to search the internet for multi-media content that they thought would help them better understand the text.  We used a procedural document that we believed might challenge many people  as the case study for our investigation. We noted a difference in both the number of searches and the types of media retrieved based on the people's familiarity with the topic. From this, we hypothesized that experts would prefer less multi-media content and novices would prefer more, particularly with respect to video content. This led us to design a second experiment that specifically analyzed the preferences of novice and expert users with respect to specific types of document augmentation.   

In the HCI community, research on augmenting reading interfaces has been ongoing to extend people's cognition and provide fluid experiences, such as by adapting forms and contents based on user needs~\cite{chang1998negotiation, egan1989formative,norman2013design}. However, less work has studied and compared how people consume multi-media features in digital documents and reading performance based on people's familiarity with the document content (e.g., novices and experts). Our work takes a fine-grained approach to study the differences in multi-media document consumption in novices and experts where we consider the in-line text, images, animated gifs, and videos as augmented content. Based on our formative study our hypotheses were 1) novices will prefer more media augmentation, particularly videos, and 2) multi-media augmentation will decrease novices' quiz time, increase their understanding, and reduce their cognitive load. 

To test these hypotheses, we implemented two versions of a digital procedural document about a titration experiment: \emph{text} focused version and media \emph{rich} version. The \emph{text} version was designed to meet our hypothesis about expert preferences, including only plain text and in-line text resolution for keywords. The media \emph{rich} version was designed to meet our hypothesis for novice preferences and it included multi-media augmentation. We recruited 48 participants and presented them with the two prototypes in a nested between-subject study to evaluate the usability of tailored prototypes, track quiz time, and understand people's multi-media features consumption patterns. We used surveys and interviews to collect data. Our results showed that, first, media preferences depend on people's familiarity with the document content. Second, multi-media content improves both consumption time and comprehension for people unfamiliar with the topic. Finally, multi-media inclusion is unnecessary for people already familiar with the topic, but it does not negatively impact reading metrics for this group. The major contributions of this work are the study and analysis of how people consume multi-media features in a digital procedural document and evaluation and analysis of two customized document experiences for two different types of users. 



\section{Related Work}

\textbf{Augmented Reading Interfaces.} In the HCI community, foundational augmentation for reading interfaces aimed to provide a "fluid" experience and customized experiences to enhance people's cognition, such as with interactive books~\cite{egan1989formative,norman2013design}. These HCI-introduced prototypes support crowdsourced answers to better explain the content in web pages~\cite{chilana2012lemonaid} or provide previews (e.g., Wikipedia preview features) to help people to jump through the content easily for navigational affordances~\cite{graham1999reader,schilit1998beyond} and to facilitate skim-reading. 
Some of the reading interfaces specifically focused on augmenting plain text with hypertext glossaries~\cite{zellweger1998fluid,conklin1987hypertext} or with social annotations~\cite{hill1992edit}. For example, these approaches were applied in scientific reading, providing in-text definitions for nonce words~\cite{head2021augmenting} and localized context~\cite{rachatasumrit2022citeread}. This can reduce their loads for searching materials on the internet, which can often return inconsistent or irrelevant results. This design was employed in a digital document reading scenario, Amazon Kindles, that shows definitions for tricky words in space between consecutive lines~\cite{kindle}. Although our work was inspired from these works (e.g., provide contextualized definitions~\cite{head2021augmenting}), these works mostly focused on text augmentation, and less work has studied augmenting reading interfaces with different types of media, such as images, animations, and videos.

\textbf{Learning Differences between Novices and Experts.} Prior work has studied how learning happens among novices and experts and their different learning requirements~\cite{patel2018analysis,collins2002third,castles2018ending,national2000people}. People's familiarity with the content impacts their engagement level~\cite{nam2020using, o2016investigating} and comprehension~\cite{mcnamara1996learning, ozuru2009prior}. For example, people unfamiliar with the content likely learn better when a document had a coherent flow, whereas people familiar with the content were less impacted by coherence. Because novices and experts have different learning requirements, this even influences people's web searching behavior as well. For example, readers who are versed in web search and are subject domain experts were more successful in web searches~\cite{lazonder2000differences,holscher2000web} than those who were not. 

To reduce the load of searching, there has been trend in digital documents including multi-media content, such as newspapers with videos. But because novices and experts have different learning requirements it can impact how people comprehend this multi-media content~\cite {chiu2020does,candello2013multimedia}. Prior work has also studied how learner expertise affects the effectiveness of multi-media content~\cite{kalyuga2013effects, chiu2017learner,kalyuga2014expertise}. These studies showed that multi-media presentations that are effective for low-knowledge learners lose their effectiveness for knowledgeable learners (expertise reversal effect). Without careful design considerations, the inclusion of multi-media content in future documents could actually detract from the reading experience~\cite{tversky2002animation}. However, it has not been well studied, if people desire multi-media content and how they consume them when reading the digital document. Thus, we should study how people differentially consume multi-media based on people's familiarity with the content and consider this as a factor when designing reading interfaces to customize people's experiences. This type of customization can benefit users, which can be part of transitioning novices to experts regarding the subject matter effectively with UI support~\cite{cockburn2014supporting}.

In summary, there has been extensive work in designing augmented documents for a general audience and in studying the differences in how novices and experts read text documents, but less work has studied how novices and experts differently consume augmented multi-media, in particular video. We were inspired by Leake et al's work~\cite{leake2020generating}, in which they studied general multi-media content preferences and developed a system that auto-generates slide shows from texts. It also goes beyond researching general preferences to specific differences in multi-media consumption between expert and novice users. For prototyping, the most similar to ours is ScholarPhi~\cite{head2021augmenting} because it aimed at augmenting a reading interface with in-line text definitions. Our work expands on this scope by further augmenting the reading interface with different richness levels of media content (images, animated gifs, and videos).

\section{Formative Study}

\begin{figure*}[h]
  \centering
  \includegraphics[width=0.7\textwidth]{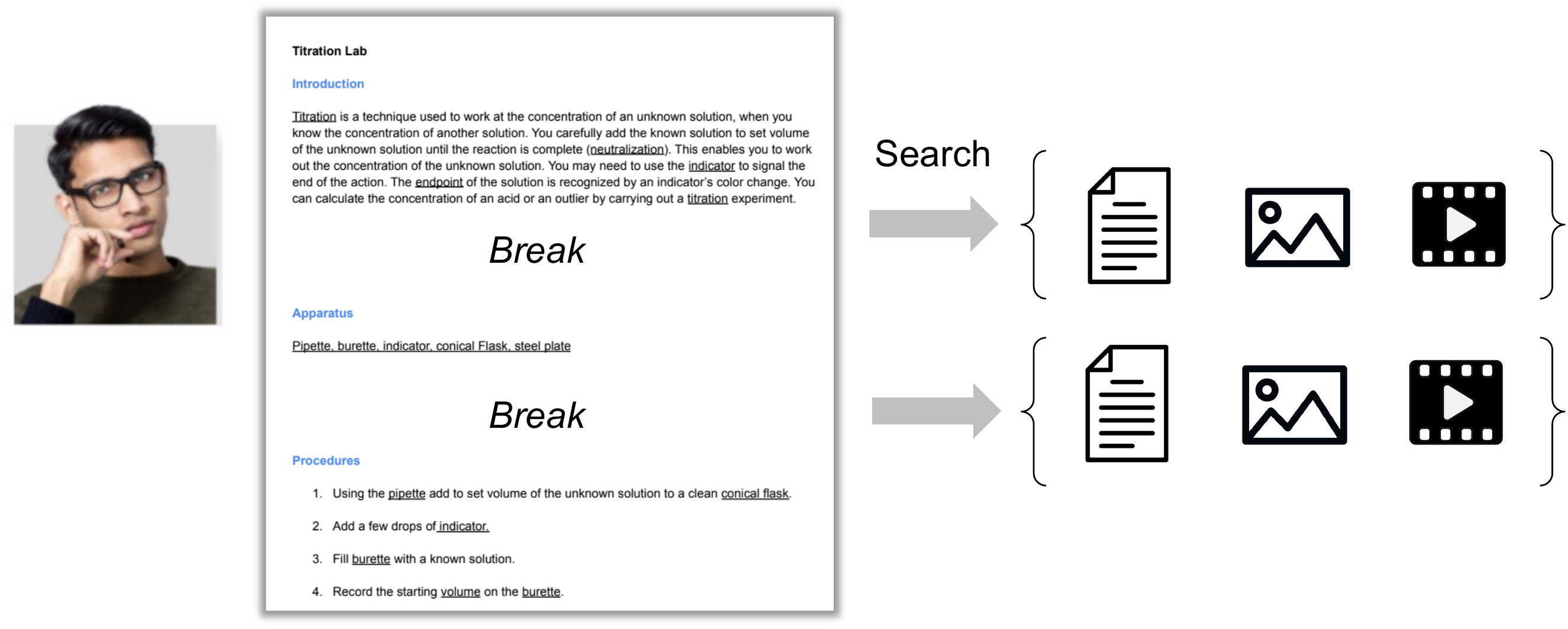}
  \caption{In our formative study, we conducted an observational study that aimed to understand user preference for media types (e.g., text, images, videos), when reading procedural documents.}
    \label{fig:formative-study-design}
    \Description{A man reading a document is given break after each paragraph to look for additional materials (e.g., text, image, video) to better understand the document. }
\end{figure*}

We wanted to understand how and why people would choose to read documents with more or less media content. To discover this, we conducted a formative study with ten participants recruited from a convenience sample. To better understand real user preferences around media consumption and reading behaviors related to mixed mode documents, we designed an experiment where participants were presented with a text document where each section was followed by a break where participants could search for additional text, images, and video content to aid their understanding (visualized in Figure ~\ref{fig:formative-study-design}). We used a procedural document that we believed might challenge many people, a chemistry titration lab, as the case study for our investigation. 

The participants were all graduate students, with backgrounds that included, research, development, and design. Participants were recruited via flyers and word-of-mouth. Before beginning, we obtained participants' consent to record the session. We introduced the study as an investigative study of how the internet can help them better understand documents.  Participants were all familiar with internet searches. At the beginning of the session, to motivate participants to do their best to understand the material we informed them that there would be a quiz at the end that they would have to pass. We told participants that they should take as much time as they needed and search for as much material as they needed to fully understand the document. We asked participants to ``think aloud" as they read and searched. During the study, we noted their comments, how often they searched, and what types of media they found. After they finished reading, we presented them with the comprehension quiz and one question regarding their satisfaction with their search results. We followed this question with a semi-structured interview. In the interview, we asked relatively open-ended provocation questions focused around: each participant's reading experience with the text-only document, their search experience, and their preferences for different types of media.

\subsection{Key Insights}
During the interviews, we asked participants about both their reading experience and their searching experience. When participants spoke about the document, one key theme that emerged was how familiar they were with the material. We began to note participants' reading speed, the types of content they searched for, and how easy it was for them to find the information they needed seemed to depend on how well they already knew the material in the document. As this insight emerged, based on participants' comments, but not on any specific evaluation, we coded our insights in terms of whether or not we believed the participant to be ``familiar" with the content or ``unfamiliar." For example, if the participants stated that they were chemistry majors or stated that they have conducted titration experiment multiples times, we coded them as \emph{familiar}. Whereas, if participants stated that they majored in liberal arts and stated that they haven't done titration experiments before, we coded them as \emph{unfamiliar}. We were able to clearly identify three of the participants as familiar with (P3, P7, P8) and three of the participants as unfamiliar (P1, P6, P10) with the material. For the remaining participants, we were not able to make a clear determination (P2, P4, P5, P9), as they did not clearly state their familiarity with the content.  We refer to our identified participants as the \emph{familiar} and \emph{unfamiliar} groups.

\textbf{Time spent on document.} \textbf{\emph{Familiar}} readers were more likely to skim through the document as they knew most of the concepts already. P3 one of the participants who skim-read the document stated that \emph{``To give you some background, I majored in chemistry when I was an undergraduate. I have done titration experiments multiple times, so I am very familiar with titrations"}. \textbf{\emph{Unfamiliar}} readers spent both more time reading the document and more time searching for unfamiliar terms on the internet, as they were learning about the materials as they read. P7 stated that \emph{``I am really not familiar with chemistry so this will be an interesting document. [...] I don't know what [word] this means, I am going to look it up. Actually, I don't know if I am looking at the right [internet search] results. [...] Because I don't really know these [apparatus] things, I had to look them up which took me longer to read. Let me know if I am taking too much time."} 

\textbf{Satisfaction with internet search results.} \textbf{\emph{Unfamiliar}} group were less satisfied with their internet search results, primarily because they had more difficulty using the right search terms to find the content they wanted. P1 stated that \emph{``I have a design background. [...] Because I am not familiar with the terms, I did not know what to look for. [...] Also, internet search results were confusing because it uses different terms interchangeably and I get confused if this refers to that or if they are the same thing."}
\textbf{\emph{Familiar}} readers were more satisfied with their internet search results because they were more often only trying to recall particular information with which they were familiar and not trying to learn concepts from scratch. P8 stated that \emph{``I did a lot of chemistry in undergrad, so I am already familiar with this kind of experiment. [...] I had to look up a few terms and apparatus to recall what they were and how to use them. [...] A quick formula look-up and definitions search was helpful."}

\textbf{Desired media augmentation.} \textbf{\emph{Familiar}} readers preferred texts over richer multi-media content such as videos. Because with texts they could skim through contents quickly by using their familiar structure. They tended not to search for videos or watch them because they perceived it as too much effort to search through and watch videos.  P7 stated that \emph{``I would not need a video for this. watching them will take too much time and I did not want to stop reading the document or get interrupted while reading. Reading is better or text is better because I can look at it when I am unsure. [...] It would be good if the definitions or formulas are supported in the document [as text]"}. In contrast, \textbf{\emph{unfamiliar}} readers desired and searched for multi-media content (images and videos) because they wanted to get a holistic picture of the document content and learn more before reading more text information. P10 stated that \emph{``I mean, video is usually always the most helpful. Because for this specifically, the video I watched at the end, it covered everything that I had looked up previously. So it was really helpful. [...] And more importantly, the video explained sequential steps, which was important for accurately performing each step, and understanding sub-steps as well, which was not covered in the text document."}

\subsection{Design considerations based on user preferences}
\label{section:design implication}
In this section, we report on user needs that were found both across all participants and some needs that were specific to more \emph{unfamiliar} readers. We call out these user needs in conjunction with design considerations that we refer to as D1, D2, D3, and D4.

\begin{itemize}
\item{D1. \bf Identify key words.} 6 of 10 participants expressed a desire to know which words represented items or concepts that were critical to the task. When asked how they would like to see important words called out, some participants suggested highlighting, underlining, boldface, color coding, or hovering over information. P6 stated that \emph{``Because I don't know these, I did not know what to search for. Because when I was reading the document, I did not want to search everything as I did not have full context about titration. [...] It will be good to denote keywords or important words I need to understand. Then I can try to understand those first"}.\\

\item{D2. \bf Provide contextualized definitions.} 6 of 10 readers from all groups expressed a desire to have definitions of keywords that were curated to the context (e.g. stir plate as it is used in titration), similar to what was done in ~\cite{head2021augmenting}. This was particularly important to the \emph{unfamiliar} group not only because it would provide relevant information right away, but also because without context these participants didn't have the knowledge to write appropriate search terms and retrieve relevant content quickly. P1 stated, \emph{``It will be good if I could see contextualized search results. When I searched things on the internet the search results were too general. I will look at those images, but I did not have the context to determine if I am seeing the right things. I felt like I needed some guidance. [...] Also, some of the results were explained differently with different terms, which made me confused."} Participants from the \emph{familiar} group also desired in-line text definitions to recall concepts quickly, so they don't have to bother with searching. P7 stated that \emph{``While reading when I encountered words that I forgot about, it's kind of like fill in the blank. I would search for that term to fill in the blank. It will be good if the document provided definitions that were curated to this experiment so that I don't have to look them up"}.\\

\item{D3. \bf Make relevant media easy to include or skip.} Adding the right level of multi-media content can provide better reading experiences~\cite{takacs2015benefits} by keeping people engaged and aiding concept recall. We found that multi-media content often helped \emph{unfamiliar} group. All of the participants searched for images and videos as well as text. P10 stated that \emph{``Because I was a beginner I did not much about the experiment and things seemed to get technical very quickly. Especially for the procedures, videos were helpful, like understanding the consequence of a certain step. For example, for the endpoint, the color changes immediately with one drop of the indicator. Being able to see that visual goal was helpful in understanding the experiment.[...] I wanted to go back and forth with a video and texts, watch a video to understand procedures, and text to skim and recall what was said in the video. So if the videos are segmented that would be good and I watch videos always on 2x speed to go through them quickly. Then the images were good for looking at specific apparatus"}.

They said, images helped them know what the apparatus looked like and videos helped them better understand how to actually conduct the experiment. By contrast, \emph{familiar} group almost exclusively searched for text information. While they would look at images that surfaced as a result of their text queries, they did not view any videos. \emph{Familiar} group stated that they preferred text as it enabled them to quickly go through the content. When asked specifically about alternative media types such as animated gifs, all \emph{unfamiliar} group expressed a desire for animated gifs showing how to do each step whereas only a minority of \emph{familiar} group said that they would like animated gifs in general but not for this content. P4 stated that \emph{``Animated gifs for each step would be helpful for following steps quickly, as it does not interrupt the reading but its kind of like a quick preview for each step."} \\

\item{D4. \bf Support different reading patterns.} The inclusion of multi-media is often distracting~\cite{takacs2015benefits,aagaard2019multitasking}, and \emph{familiar} group did not find it valuable. Because \emph{familiar} group mentioned that they skimmed the document, we hypothesized that excluding multi-media content would decrease distraction and create a more skim-friendly reading experience for this type of people. 
\end{itemize}

\section{Designing Media Augmented Documents for Novices and Experts}
\label{section:user interface}

Based on the findings of the formative study, we wanted to further investigate designing media-augmented documents specifically for people who are either familiar or unfamiliar with the topic. We will now refer to \emph{familiar} readers as \emph{experts} and \emph{unfamiliar} readers as \emph{novices}. To this end, we developed two different prototypes, a \emph{text} focused document for \emph{experts} and a media \emph{rich} document for \emph{novices} following the design insights outlined in Section~\ref{section:design implication}.

\subsection{Design Process} Our prototype development followed an iterative design process. To refine our design, we repeated the process of ideation, prototyping, and getting feedback in several rounds. Researchers met weekly to ideate early prototype design, then conducted informal usability testing to get feedback every two to three weeks. In these meetings, researchers discussed what multi-media features to support and when and how to support them in digital documents. We actively recruited potential users with various levels of expertise to get different perspectives. Design alternatives for augmented reading interfaces were identified through literature review (including hypertexts~\cite{zellweger1998fluid}, summarization~\cite{allahyari2017text}, and expansions) and evaluation of commercial tools (including Youtube's video timestamp and Wikipedia's preview).

\subsection{Stimulus Document}


We used the chemistry titration lab as our stimulus document to build our differently augmented multi-media documents. We use this scenario as a case study. Knowing that it might be difficult to find appropriate media with which to augment a text document, the text chemistry titration lab was derived from content originally in a video format. The video explains how to carry out a 10-step titration experiment including the goal of the experiment, the apparatus needed, and the procedural steps. We transcribed the video to create the text-only version. We then had the original video from which to derive the other multi-media content including images, animated gifs, and videos with timestamps. To make the two prototypes as comparable as possible, we kept the text identical but added different media types to the \emph{novice} version. We implemented these prototypes using Adobe XD and with  Anima~\cite{Anima} plugin.

\textbf{Expert version.} We designed an expert version as a text-focused document. Following the design insight outlined in section~\ref{section:design implication} we included \emph{keyword underlining} and \emph{in-line text augmentation} to facilitate skim reading and quick recall of key facts. Keywords were frequently used important words in the titration experiment. These words were underlined and users could retrieve the tooltip of a text definition (in-line text) by hovering over the underlined word.

\textbf{Novice version.} We designed a second prototype for \emph{novice} group augmented with \emph{rich} media content, following our design insights. This version augments the \emph{expert} version with \emph{keyword underlining \& in-line text argumentation, key images describing the apparatus, animated gifs describing each step of the procedure, and a time-stamped bookmarked video}.

\section{Study Design}

We wanted to evaluate if our customized digital document provided the right level of media content for each of our identified groups: \emph{novice} and \emph{experts}. We designed a task in which each group read a document and then took a quiz to understand how different groups consume multi-media features and measure their comprehension. We allowed people to refer to the document during the quiz. Our hypotheses for the experiment were:

\begin{itemize}
    \item \textbf{(H1):} The \emph{expert} group will prefer \emph{text} version and \emph{novice} group will prefer the media \emph{rich} version.
    \item \textbf{(H2):} The \emph{expert} group will read faster with \emph{text} version.
    \item \textbf{(H3):} The \emph{novice} group will spend more time reading the \emph{rich} version but will complete the quiz faster, score higher on a quiz, and experience reduced cognitive load.
\end{itemize}




Our study design was a two-way between-subjects study for each version of the document. We recruited separate cohorts of \emph{experts} and \emph{novices} using a screener survey, to experience either the media \emph{rich} document or the \emph{text} document. In total, we recruited 48 participants (24 experts and 24 novices). Within each cohort, participants were randomly assigned to either the \emph{text} or media \emph{rich} version of the document.

\textbf{Participants.} We recruited 48 participants (female: 22, male: 26) from diverse backgrounds that included engineering, education, chemistry, product management, and design. For moderated sessions, the study took 40 minutes to complete and we compensated participants with a \$35 gift card. For unmoderated sessions, the study took 20 minutes and we compensated participants with a \$10 gift card. The moderated sessions were identical to the unmoderated sessions except that the instructions were given verbally by the moderator and the session was followed by a semi-structured interview. For the unmoderated version, participants read from the instructions.  For moderated sessions, users were recruited via word of mouth and flyers and for unmoderated sessions, they were recruited via User Testing~\cite{usertesting} platform. To identify novices and experts, we used a screener that asked potential participants about their backgrounds and areas of expertise. To qualify as experts, participants needed to answer a majority of these questions (experts above 70 \%, novices below 50 \%, out of 10 questions) correctly.

\textbf{Procedure. }For qualified participants, we collect consent for the study. All participants were informed that they should take as much time as they needed to read the document to fully understand it.
In the beginning, they were told there would be a comprehension quiz given after they finished reading, but that they would be allowed to refer to the document during the quiz. Participants were then allowed to read the document naturally, without further guidance (unguided reading). The full versions of each prototype are shown in Figure~\ref{fig:teaser}. The comprehension quiz was given after each participant finished reading, followed by a survey that collected quantitative measures of the usability (SUS)~\cite{brooke1996quick} and cognitive load~\cite{hart2006nasa}. The majority of the experiments were unmoderated, using a remote testing platform (User Testing~\cite{usertesting}). We randomly selected two participants in each condition, eight total, to participate in a moderated version of the study.

\section{Results}
In this section, we report on both objective measures of performance including reading time, quiz time, and comprehension scores as well as subjective measures including scores for usability and NASA TLX mental demand and effort level. These results are summarized in Table ~\ref{tab:user-performance} - ~\ref{tab:usability significance} and described in detail below. We additionally include qualitative feedback from the semi-structured interviews of the moderated sessions.

\subsection{Objective measures.} We tracked reading time and quiz completion time from our recorded screen session videos. We also noted whether or not participants watched videos in the media \emph{rich} document condition. T-test results are summarized in Table~\ref{tab:user-performance} and ~\ref{tab:user perfomance significance}.  

\begin{table*}[t]
  \caption{Objective Measures}
  \label{tab:user-performance}
  \begin{tabular}{c|c|c|c}
    \toprule
    Conditions & Reading Time (seconds) & Quiz Time (seconds) & Quiz score\\
    \midrule
    Novice,rich & M = 319.58, SD = 101.08 & M = 149.99, SD = 45.92 & M = 8.66, SD = 1.15 \\
    Novice, text & M = 213.58, SD = 137.54 & M = 299.15, SD = 191.82 & M = 6.910, SD = 2.02 \\
    Expert,rich & M = 251.41, SD = 121.86 & M = 187.53, SD = 118.85 & M = 8.83, SD = 0.71 \\
    Expert,text & M = 175.41, SD = 79.36 & M = 164.97, SD = 92.02 & M = 8.58, SD = 0.66 \\
  \bottomrule
\end{tabular}
\end{table*}

\begin{table*}[t]
  \caption{T-test Objective Measures Analysis. * denotes significance}
  \label{tab:user perfomance significance}
  \begin{tabular}{c|c|c|c}
    \toprule
    Compare & Reading Time & Quiz Time & Quiz Score\\
    \midrule
    Novice,rich \& Novice,text & \textbf{*t(22) = 2.15 , p = .04} & \textbf{*t(22) = -2.61 , p = .01} & \textbf{*t(22) = 2.60 , p = .01} \\
    Expert,rich \& Expert,text & t(22) = 1.81 , p = .08 & t(22) = 0.51 , p = .61 & t(22) = 0.42 , p = .67 \\
    
    
    Expert, text \& Novice,text & t(22) = -0.83, p = .41 & \textbf{*t(22) = -2.18, p = .03} & \textbf{*t(22) = 2.71, p = .01}\\
    
    Novice, rich \& Expert, rich & t(22) = 1.49, p = .15  & t(22) = -1.02, p = .31 & t(22) = -0.42, p = .67\\
  \bottomrule
\end{tabular}
\end{table*}

\textbf{Reading Time.} We found significant differences in reading time, for \emph{novices} between the media \emph{rich} (M = 319.58s, SD = 101.08) document and the \emph{text} (M = 213.58s, SD = 137.54) document, (t(22) = 2.15, p = .04). But we did not find this significant difference for \emph{experts} across conditions, (t(22) = 1.81, p = .08). This means that \emph{novices} in media \emph{rich} condition spent significantly longer time and consumed multi-media features in the document compared to \emph{novices in text} condition. 

We noted that \emph{novices} more frequently choose to watch the video from the beginning to the end (10 out of 12 participants) in the media \emph{rich} document and all \emph{novices} played the video at normal speed. Only half (6 out of 12 participants) of \emph{experts} chose to watch a video in the media \emph{rich} document and the majority of \emph{experts} watched only a portion of the video.

\textbf{Quiz time}. We similarly found significant differences in quiz time for \emph{novices} between the media \emph{rich} document (M = 149.99s, SD = 45.92) and the \emph{text} document (M = 299.15s, SD = 191.82), (t(22) = -2.61, p=.01). But we did not find a significant difference for \emph{experts} across the conditions, (t(22) = 0.51, p = .61). This means that \emph{novices} benefited from the media \emph{rich} document as they were able to finish quizzes on time faster than \emph{novices} in \emph{text} condition. On the other hand, \emph{experts} did not significantly benefit from multi-media content. 

We also found a significant difference between \emph{experts} and \emph{novices} with respect to completing quiz for the \emph{text} document with \emph{experts} spending (M = 164.97s, SD = 92.02) and \emph{novices} spending (M = 299.15s, SD = 191.82), (t(22) = -2.18, p = .03). This means that \emph{novices in text} condition spent a significantly longer time in finishing a quiz than \emph{experts in text} condition.

\textbf{Quiz score.} We again found significant differences for \emph{novices} between the \emph{text} document (M = 6.91, SD = 2.02) and the media \emph{rich} document (M = 8.66, SD = 1.15), ((t(22) = 2.60, p = .01). But we did not find significant differences for \emph{experts} across the conditions, (t(22) = 0.42, p = .67). This means that \emph{novices} in media \emph{rich} condition, who had access to multi-media content scored significantly higher on the quiz than those in \emph{novices in text} condition. In both conditions, \emph{experts} scored high on quizzes regardless of having access to multi-media content. We hypothesize that this is because \emph{experts} knew most of the content already and they were using the procedural document to recall concepts. 

We also found a significant difference between \emph{experts} and \emph{novices} with respect to quiz scores for the \emph{text} document with \emph{experts} scoring (M = 8.58, SD = 0.66) and \emph{novices} scoring (M = 6.91, SD = 2.02), (t(22) = 2.71, p = .01). This means that \emph{expert in text} condition scored significantly higher than \emph{novices in text} condition.

\subsection{Subjective Measures: Usability \& Cognitive Load.} For our subjective analysis, we asked participants to take a survey that included SUS scale questions to assess usability~\cite{brooke1996quick} and NASA-TLX questions to assess perceived mental demand and effort level~\cite{hart2006nasa} (see Table~\ref{tab:usability} and ~\ref{tab:usability significance}). 

\textbf{Usability.} Using the SUS scores, we found significant differences in usability scores for \emph{novices} between the \emph{text} document (M = 63.95 (below average), SD = 18.56) and the media \emph{rich} document (M = 83.33 (excellent), SD = 14.93), (t(22) = 2.81, p = .01). But we did not find significant differences for \emph{experts} across the condition, (t(22) = 1.52, p = .14). This means that \emph{novices} in media \emph{rich} condition reported significantly higher usability scores than \emph{novices in text} condition. 

But we did not see a significant difference in perceived usability scores across \emph{experts in media rich} and \emph{experts in text} conditions, as they largely ignored multi-media features and were satisfied with in-line text augmentation. 
We also found that reported usability scores were higher for \emph{experts in text} condition (M = 75.83 (good), SD = 16.69) than \emph{novices in text} (M = 63.95 (below average), SD = 18.56) condition, (t(22)=2.20,p = .03). 

\textbf{Mental demand \& workload.} We did not find a significant difference in participants' perceived mental demand and effort level across the conditions.



\begin{table*}[t]
  \caption{Subjective Measures}
  \label{tab:usability}
  \begin{tabular}{c|c|c|c}
    \toprule
    Conditions & Usability Score & Mental Demand & Effort Level\\
    \midrule
    Novice,rich & M = 83.33, SD = 14.93 & M = 3.5, SD = 1.80 & M = 3.5, SD = 1.83 \\
    Novice, text & M = 63.95, SD = 18.56 & M = 4.27, SD = 2.24 & M = 3.91, SD = 2.02 \\
    Expert,rich & M = 87.5, SD = 15.73 & M = 3.18, SD = 1.40 & M = 2.83, SD = 1.11 \\
    Expert,text & M = 75.83, SD = 16.69 & M = 3.09, SD = 1.57 & M = 2.7, SD = 0.94 \\
  \bottomrule
\end{tabular}
\end{table*}

\begin{table*}[t]
  \caption{T-test Subjective Measures Analysis. * denotes significance}
  \label{tab:usability significance}
  \begin{tabular}{c|c|c|c}
    \toprule
    Compare & Usability Score & Mental Demand & Effort Level\\
    \midrule
    Novice,rich \& Novice,text & \textbf{*t(22) = 2.81 , p = .01} & t(22) = -0.83 , p = .41 & t(22) = -0.52 , p = .60 \\
    
    Expert,rich \& Expert,text & t(22) = 1.52 , p = .14 & t(22) = 0.11 , p = .91 & t(22) = -0.61 , p = .54 \\
    
    Expert, text \& Novice,text & \textbf{*t(22) = 2.20, p = .03} & (22) = -1.01, p = .32 & t(22) = -1.02, p = .31\\
    
    Novice, rich \& Expert, rich  & t(22) = -0.66 , p = .51 & t(22) = 0.11 , p = .90 & t(22) = 1.07 , p = .29 \\

  \bottomrule
\end{tabular}
\end{table*}


\subsection{Qualitative Feedback}
We gathered qualitative feedback from the eight people who participated in our moderated study. This included two participants from each of the four assigned conditions, referred to as \emph{novice, rich} as NR, \emph{novice, text} as NT, \emph{expert, rich} as ER, and \emph{expert,text} as ET.

\textbf{Novices.} \emph{Novices} that interacted with \textbf{\emph{text}} version (NT1-2) desired richer media content such as images and videos that explain what the apparatus look like. NT2 stated \emph{``I would have liked an image or a video [in addition to text] that explains each step in more detail. [...] I think it was a bit confusing for me to differentiate the apparatuses used in the experiment. But I did find the tooltip definition helpful and easy to use."} 
\emph{Novices} that interacted with \textbf{\emph{rich}} version found the multi-media content helpful for understanding document content (NR1-2). NR1 said \emph{``I found the videos helpful to learn steps about how to conduct the titration experiment. If you noticed, when I was taking the quiz I was going back to the animated gifs to recall some of the contents that needed clarification."} NR2 stated that \emph{``As I did not know about this material, it was helpful to see these materials visually. Especially, I liked the fact that the video and animated gifs were explaining the materials I am reading exactly. [...] While reading, if I get lost, I can go watch the video or look at animated gifs to proceed. It was good to have those information [visual representation] upfront so that I did not have to look up those materials on my own."}\\

\textbf{Experts.} \emph{Experts} that interacted with \textbf{\emph{text}} version mentioned that they had all the information they needed and did not desire an additional layer of media content (ET1-2). They also appreciated in-line text augmentation. ET1 stated \emph{``I am an industrial chemistry major student, so I know how to carry out a titration experiment, which I did multiple times. [...] This document was enough for me and had all the information I needed.[...] I don't wish to watch a video or something"}. ET2 stated, \emph{``I am familiar with titration, but sometimes I had to recall some of the concepts. And there are different types of titration experiments, like acid-based titration and others. [...] I liked the tooltip that had definitions that explained what I was looking for immediately, which was what I needed. And those explanations were directly related to the context [explanations were customized to the specific type of experiment]"}. \emph{Experts} that interacted with the \textbf{\emph{rich}} document did not seem to attend to the multi-media content (ER1-2). ER1 stated \emph{``I did think about watching the video because it was there. But I realized the same information was in the text and it was just easier to read. I may have looked at the animated gifs because it was there but I mostly read text and used tooltip when I get stuck in the document.[...] If I were to watch the video, I would have to look for a place [video segment] that explains it and that was too much. Also, if I start watching the video that interrupts the reading, you know, and I did not want to stop reading"}. It was observed that both \emph{experts} who interacted with the media \emph{rich} version in the moderated session seemed to skim through the document. They used only a few in-line text pop-ups to recall concepts and neither \emph{expert} in the moderated sessions watched any part of the video.
\section{Discussion}

We wanted to develop a procedural document that included the right level of media content for all users. To do that we conducted a formative study to understand how people consume a text-only procedural document, which investigated what type of media content they additionally desired. Our analysis revealed that people's familiarity with the subject matter impacted how they read the document and consume multi-media features. This led us to design a second experiment that specifically targeted \emph{novice} and \emph{expert} users. We implemented two versions of a media-augmented document: \emph{text} document and media \emph{rich} document. In our second experiment, we evaluated if these two documents provided the right level of media content for our identified groups: \emph{novices} and \emph{experts}. Our results showed that:

\begin{itemize}
\item H1:(true) {\em Expert group will prefer the text version and novice group will prefer media rich version.} In general, \emph{experts} preferred text (augmented) as it met all of their information needs. They largely ignored the multi-media content in the media \emph{rich} version as evidenced by the qualitative results of the moderated interviews.

\item H2: (not true) {\em The expert group will read faster and complete the quiz faster with text version.}

\item H3: (mostly true) {\em The novice group will spend more time reading the media rich version but will complete the quiz faster, score higher on the quiz, and experience reduced cognitive load.}

\end{itemize}

\textbf{User preference for multi-media consumption.} Our results showed that multi-media content, in particular video, was not always desired and that it depended on the user and their familiarity with the document content. We believe that customizing the amount of multi-media content: images, animated gifs, and videos included in documents could increase productivity and decrease distraction. We envision that variable multi-media inclusion could be achieved in a number of ways. One simple method would be to provide a UI element that allows users to manually self-select media content based on their preferences. Another method would be to identify novices and experts automatically based on their reading history~\cite{nam2020using}. Given that many websites and e-readers collect information regarding user behaviors, we hypothesized that a user's familiarity with a given topic could be inferred from this collected data, then websites could have a plug-in feature to  support the right richness level of multi-media content, such as including more video or simply augmented text.
We imagine that our work can be applied to alternative contexts and scenarios, such as customizing the level of multi-media support based on learner expertise in Education~\cite{chiu2020does}, such as chemistry, medicine, vocabulary learning~\cite{balslev2005comparison, cocco2021comparing,arndt2018vocabulary} and for experts and novices visitors in tourism at cultural heritage~\cite{candello2013multimedia}.


\textbf{Dynamic multi-media augmentation.} We also believe that a user's familiarity with the content can be dynamically assessed as the reader is consuming the document based on reading speed or other behaviors by using predictive statistical models~\cite{zukerman2001predictive,frias2006survey}.
In our study, we presented two fixed content documents either with or without multi-media content, however, this content could be included dynamically based on real-time user behaviors. For example, when reading speed slows the document could transform text to images~\cite{zhu2007text}, text to videos to better explain content~\cite{zhu2007text}, or algorithms could even transform the entire document into an audio-visual slide show~\cite{leake2020generating}. Changing the amount of multi-media content dynamically supports automatic customization making documents resilient to changing user preferences, such as being able to always add content (expansion of information).




\section{Conclusion}


We had the high-level goal of developing digital documents with the right level of media augmentation for all users. To this end, we conducted an observational study with real users reading a procedural document. Our study revealed that depending on people's familiarity with the content, people deferentially consumed multi-media features in digital documents. To confirm our hypothesis that the desire for multi-media content was dependent on a people's familiarity with the material, we next conducted a second user study with 48 different participants where we specifically recruited \emph{novices} and \emph{experts} in chemistry and had them read either a \emph{text} focused or media \emph{rich} presentation of a chemistry titration lab document. Our findings showed that experts preferred text and novices benefited from multi-media content. That said, customizing the richness of multimedia in documents for different types of users could increase productivity and decrease distraction.

\bibliographystyle{ACM-Reference-Format}
\bibliography{sample-base}

\appendix

\end{document}